# Dynamic Risk Assessment for Geologic CO$_2$ Sequestration

Bailian Chen[1], Dylan R. Harp[1], Yingqi Zhang[2], Curtis M. Oldenburg[2], and Rajesh J. Pawar[1]

1. *Earth and Environmental Sciences Division, Los Alamos National Laboratory, Los Alamos, NM 87544*

2. *Energy Geosciences Division, Lawrence Berkeley National Laboratory, Berkeley, CA 94720*

**Abstract:** At a geologic CO$_2$ sequestration (GCS) site, geologic uncertainty usually leads to large uncertainty in the predictions of properties that influence metrics for leakage risk assessment, such as CO$_2$ saturations and pressures in potentially leaky wellbores, CO$_2$/brine leakage rates, and leakage consequences such as changes in drinking water quality in groundwater aquifers. The large uncertainty in these risk-related system properties and risk metrics can lead to over-conservative risk management decisions to ensure safe operations of GCS sites. The objective of this work is to develop a novel approach based on dynamic risk assessment to effectively reduce the uncertainty in the predicted risk-related system properties and risk metrics. We demonstrate our framework for dynamic risk assessment on two case studies: a 3D synthetic example and a synthetic field example based on the Rock Springs Uplift (RSU) storage site in Wyoming, USA. Results show that the NRAP-Open-IAM risk assessment tool coupled with a conformance evaluation can be used to effectively quantify and reduce the uncertainty in the predictions of risk-related system properties and risk metrics in GCS.

# 1. Introduction

One of the main concerns with geologic $CO_2$ sequestration (GCS) projects is the potential risks of $CO_2$ and brine leakage to overlying resources (e.g., underground sources of drinking water (USDW), hydrocarbon and mineral resources) (Benson and Myer, 2003; Harp et al., 2016; Xiao et al., 2020). To build confidence of stakeholder, a scientific approach to quantitatively manage risk is needed to provide accurate predictions of long-term risks of $CO_2$ sequestration systems (Condor et al., 2011; Pawar et al., 2013; De Lary et al., 2015; Li and Liu, 2016; Pawar et al., 2016).

Numerous studies have demonstrated application of quantitative approaches for risk assessment. Stauffer et al. (2009) developed $CO_2$- Predicting Engineered Natural Systems ($CO_2$-PENS) for GCS performance assessment and risk analysis. The design of $CO_2$-PENS is for the purpose of performing probabilistic simulations of $CO_2$ capture, transport, storage, and leakage to overlying aquifers and ultimately the atmosphere. Zhang et al. (2011) developed a $CO_2$ sequestration module on the basis of the CQUESTRA (Carbon dioxide seQUESTRAtion) model for probabilistic risk assessment. They showed that significant $CO_2$ leakage is not likely for a site with a single injection well, while multiple potentially leaky wells present the risk of measurable leakage. Nicot et al. (2013) leveraged the certification framework (CF) to assess the risks of $CO_2$ and brine leakage from a storage reservoir to various overlying components such as USDWs and near-surface environments. The utilization of the CF approach to the Southeast Regional Carbon Sequestration Partnership (SECARB) Phase III $CO_2$ injection site indicated that the risks for $CO_2$ and brine leakage are both low. The U.S. Department of Energy's National Risk Assessment Partnership (NRAP) developed a scientific prediction tool for risk assessment called the Integrated Assessment Model for Carbon Sequestration (NRAP-IAM-CS) (Pawar et al., 2013; Pawar et al., 2016), and $CO_2$-PENS model was utilized as a foundation for developing this tool. The NRAP-IAM-CS separates a GCS operation into its key components (e.g., geologic reservoir, leakage pathway, groundwater aquifers) and simulates the physical processes within each component separately. Onishi et al. (2019) applied the NRAP-IAM-CS tool to assess GCS risk for a carbonate reservoir, called Kevin Dome in Montana. They found that the potential amount of $CO_2$ leaked is affected by permeability, residual $CO_2$ saturation, $CO_2$ relative permeability hysteresis, confining rock permeability, and capillary pressure. Xiao et al. (2020) conducted risk assessment for an active $CO_2$ enhanced oil recovery (EOR) field, The Farnsworth Unit in Texas. The $CO_2$ and brine leakage risks to the overlying USDW were quantified with a proxy modeling approach. Most recently, NRAP has developed a Python-based open-source IAM (NRAP-Open-IAM) to help address questions about a potential GCS site's ability to effectively contain injected $CO_2$ and protect groundwater and other overlying environmentally sensitive receptors, and facilitate stakeholder decision-making about the safety and effectiveness of GCS. NRAP-Open-IAM has a collection of reduced order models (ROMs) for each potential system component in a $CO_2$ storage site, and a number of tools that can be used for risk assessment. A core capability of the NRAP-Open-IAM is to allow a user to execute stochastic and dynamic simulation of whole GCS system performance and leakage risk assessment very quickly (Vasylkivska et al., 2021).

The work for risk assessment mentioned above did not investigate the value of information of monitoring data (e.g., pressure, temperature and $CO_2$ saturation data) collected from monitoring wells during the operation of $CO_2$ storage. It has been demonstrated by several studies that monitoring data can contribute to reduce the uncertainty of predicted risk-related system properties and risk metrics (i.e., narrowing of uncertainty bands). Oladyshkin et al. (2013) developed a workflow using bootstrap filtering and reduced-order models (ROMs) to integrate pressure measurements into reservoir models and evaluate the reduction of uncertainty in $CO_2$ leakage rate at a sequestration site. Several uncertain parameters, namely, wellbore permeability, reservoir permeability, and reservoir porosity, were considered in their work. Chen et al. (2018) presented a methodology based on a filter-based data assimilation method and a proxy model to conduct network design of $CO_2$ monitoring. The optimal monitoring solution was chosen by reducing the uncertainty in the forecast of the total amount of $CO_2$ leakage. Although the methods proposed by Chen et al. and Oladyshkin et al. for assimilation of monitoring data are computationally efficient, their approaches are limited to situations involving only a limited set of uncertain parameters. Sun and Durlofsky (2019) proposed an approach using data-space inversion (DSI) to quantify uncertainty in the predictions of $CO_2$ plume locations. In the DSI based approach, the distributions of $CO_2$ saturation are predicted using simulation results based on prior models together with monitoring data. It is worth to mention that posterior models (updated models) conditional to observations were not generated in the DSI approach. This is different from the traditional data assimilation approaches such as the well-known ensemble-based methods (e.g., Ensemble Kalman Filter). Recently, Chen et al. (2020) demonstrated how uncertainty in the predictions of risks can be reduced by conducting assimilation of monitoring data, where the data assimilation is performed by using an advanced version of a state-of-the-art data assimilation approach, the Ensemble Smoother with Multiple Data Assimilation (ES-MDA) (Emerick and Reynolds, 2013). The risk assessment considered in their work mainly focuses on the quantification of risk-related system properties in the reservoir (e.g., pressure and $CO_2$ saturation and plume areas). In this paper, we extend the work of Chen et al. (2020) and conduct a more comprehensive dynamic risk assessment where the risk can originate not only from the reservoir but also from wellbores and groundwater aquifers.

This paper proceeds as follows: first, we present the proposed framework for dynamic risk assessment and then we describe the two major components of the framework, i.e., conformance evaluation and risk assessment. Next, we demonstrate the proposed framework for dynamic risk assessment with two case studies: a 3D synthetic example and a synthetic field example based on the RSU site in Wyoming, USA. Finally, we present the conclusions of this paper.

## 2. Methodology

### 2.1. Dynamic risk assessment

The proposed framework for dynamic risk assessment is presented in Figure 1. As can be observed from Figure 1, two main components are included in this framework, i.e., conformance evaluation and risk assessment. The conformance of a $CO_2$ sequestration system is defined as the condition under which there is acceptable past and current concordance and acceptable forecasted performance (Oldenburg, 2018). Concordance quantifies the agreement between observations and simulations, while performance indicates that the GCS operation is working to specifications, e.g., $CO_2$/brine leakage rates below acceptable thresholds. Risk assessment for GCS is the overall process of identifying, analyzing, and quantifying potential risks. Risk assessment is part of a risk management strategy utilized to quantify potential risks during the injection and post-injection phases of a GCS site. Here, dynamic risk assessment is defined as the process of iteratively identifying and evaluating GCS risks when observational data or measurements become available from a storage site. This goal of the dynamic process is to reduce the uncertainty in the predictions of risk-related system properties and risk metrics in GCS. Next, we give a brief summary of the proposed framework for dynamic risk assessment.

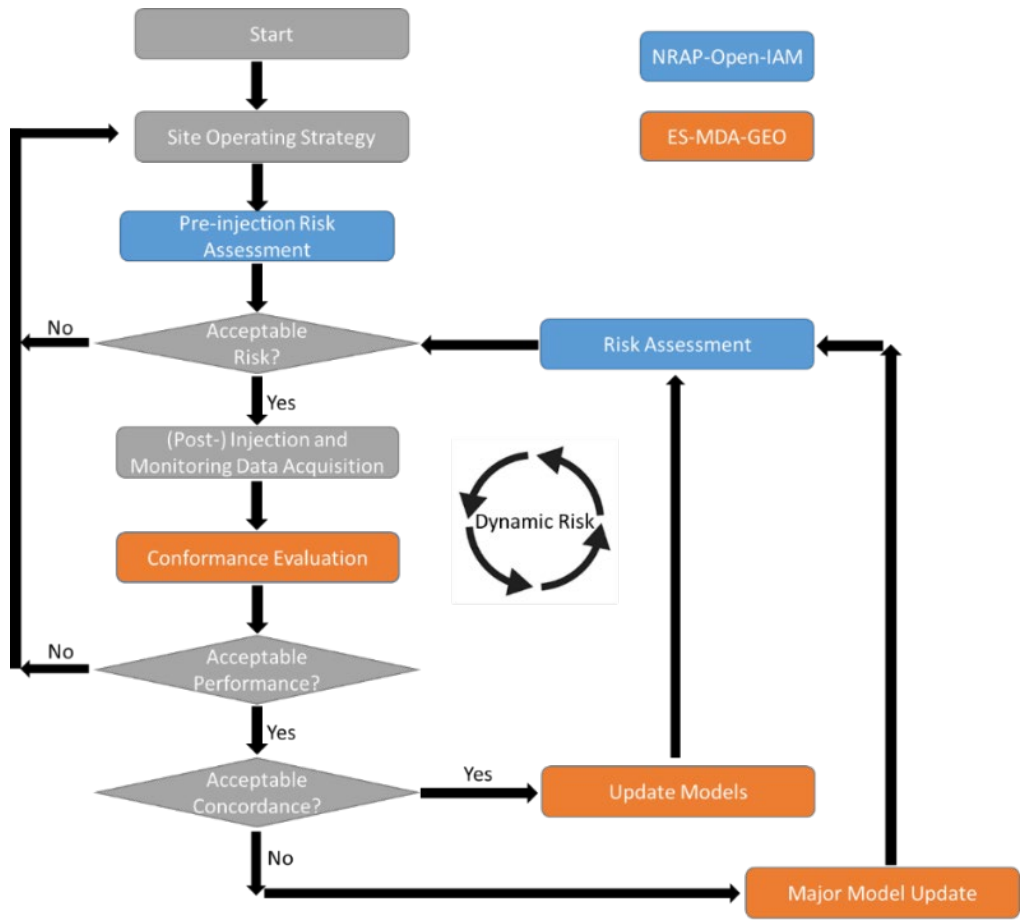

**Figure 1. The framework for dynamic risk assessment in geologic CO$_2$ sequestration.**

**Step 1.** Design storage site operating strategy. With an identified CO$_2$ storage site, the site operating strategy including the number and locations of injection/monitoring wells and CO$_2$ injection schedule will be determined. Numerous studies have been conducted to identify the optimal site operating strategy; refer to Zhang and Agarwal (2013), Yonkofski et al. (2016), Sambandam (2018), Chen et al. (2018), González-Nicolás et al. (2019), etc. Note the design of the site operating strategy is not the scope of this paper, and we assume the initial operating strategy is pre-determined.

**Step 2.** Perform pre-injection risk assessment. Prior to $CO_2$ injection, a risk assessment can be conducted using NRAP-Open-IAM or other GCS risk assessment tools to determine whether the potential risks (e.g., $CO_2$/brine leakage risks) are acceptable. If any of the potential risks are not acceptable, then one needs to go back to Step 1 and adjust the site operating strategy until all the assessed risks are within the acceptable threshold.

**Step 3.** Start $CO_2$ injection and monitoring data acquisition. After $CO_2$ injection gets started, measurements such as pressure and $CO_2$ saturation in monitoring wells will become available. These measurements are usually collected with a particular frequency. Here, we use the data collected once per month, although they may be collected at higher frequency. The collected monitoring measurements will be integrated into reservoir models for conformance evaluation in the next step.

**Step 4.** Conduct conformance evaluation and update reservoir models. In the conformance evaluation, the forecasted performance (i.e., GCS operation is working to specifications) needs to be evaluated first. If the forecasted GCS performance is acceptable, the next part of conformance can be evaluated, namely the agreement between observations (i.e., monitoring measurements) and reservoir simulation predictions. If there are minor discrepancies between observations and predictions made by the reservoir models, updates to the models will be made using ES-MDA-GEO, which is the most advanced version of the state-of-the-art data assimilation approach ES-MDA (Rafiee and Reynolds, 2017). Note that when there is a large discrepancy between observations and simulated results, a process of major model update by incorporating more measurements such as geophysical monitoring data and seismic data may be required.

**Step 5.** Re-calculate the predicted risk with the updated reservoir models. With the updated reservoir models, we can make more accurate predictions, with less uncertainty in risk-related system properties such as pressure (P) and $CO_2$ saturation (S) plumes, and P/S in monitoring and legacy wells. These improvements can also improve the accuracy of integrated assessment modeling and the predictive accuracy in the risk-related metrics or quantities, e.g., $CO_2$ and brine leakage rates and groundwater aquifer volumes with pH/TDS change. The risk assessment will be conducted periodically until no significant uncertainty reduction can be observed in the risk assessment by incorporating measurements from monitoring wells.

Next, more technical details about conformance evaluation and risk assessment will be provided.

## 2.2. Conformance evaluation

In this study, the conformance evaluation was performed using a data assimilation approach called ES-MDA with *geo*metric inflation factors (ES-MDA-GEO) (Rafiee and Reynolds, 2017). It has been demonstrated that ES-MDA is superior over EnKF for data assimilation or history matching. Although the original ES-MDA has already been shown to be an effective approach for assmilating various types of data collected from subsurface (Silva et al., 2017; Zhao et al., 2017; Emerick, 2018; Evensen, 2018; Kim et al., 2018; Zhang et al., 2020), the major drawback of the original ES-MDA algorithm is that in each data assimilation step, the inflation factors must be predetermined before the process of data assmilation. To resolve this critical issue with the application of the original ES-MDA algorithm for data assimilation, Le et al. (2016) developed an adaptive approach to iteratively update the inflation factors at each step of ES-MDA data assimilation. Although this adaptive approach has reasonably improved the performance of the original version of ES-MDA algorithm, it usually needs a large number of data assimilation steps to converge, which may be computationally prohibited for large-scale field cases. Rafiee and Reynolds (2017) developed an effective and efficient approach to compute the inflation factor used at each step for data assimilation, which allows users to specify the total number of steps to be used in the process of data assimilation based on the available computing resources, while at the same time allowing enough changes in the reservoir models at each data assimilation step to control the issue of over- and under-shooting that can lead to inaccurate inversion or update of reservoir geological models. This version of the ES-MDA algorithm is called ES-MDA-GEO and was utilized in this work for assimilating data collected from monitoring wells during sequestration operations.

Through conformance evaluation via ES-MDA-GEO-based data assimilation, temporal $CO_2$ sequestration site-monitoring data can be integrated into geological models. The uncertainty in reservoir parameters such as a heterogeneous permeability field can subsequently be reduced.

### 2.3. Risk assessment

The risk assessment for GCS was conducted using an open source Integrated Assessment Model (NRAP-Open-IAM) developed by the U.S. Department of Energy's National Risk Assessment Partnership (NRAP) (Vasylkivska et al., 2021). NRAP-Open-IAM has been developed to perform stochastic simulation of whole GCS system performance, leakage risk assessment and uncertainty quantification. NRAP-Open-IAM is more user-friendly (e.g., open source and customizable by its users) and has a number of new features and capabilities (e.g., uncertainty reduction and risk management) relative to its predecessor, NRAP-IAM-CS. There are three major component model types in NRAP-Open-IAM: reservoir, leakage pathways, and receptors (see Figure 2). Component models in NRAP-Open-IAM are coupled such that the outputs of one component provide inputs to the other component models.

By coupling different components of a GCS system in the integrated assessment modeling, the uncertainty in the predictions of different risk metrics, e.g., such as $CO_2$/brine leakage rates and pH/TDS plume size, can be effectively quantified. By combining the conformance evaluation process with the risk assessment tool NRAP-Open-IAM, we can dynamically quantify the impact of utilizing monitoring measurements on reducing uncertainty in the predictions of different risk-related system properties such as saturation/pressure in leaking wellbores, $CO_2$ and brine leakage rates from wellbores, and risk metrics for groundwater aquifer impact such as pH plume size.

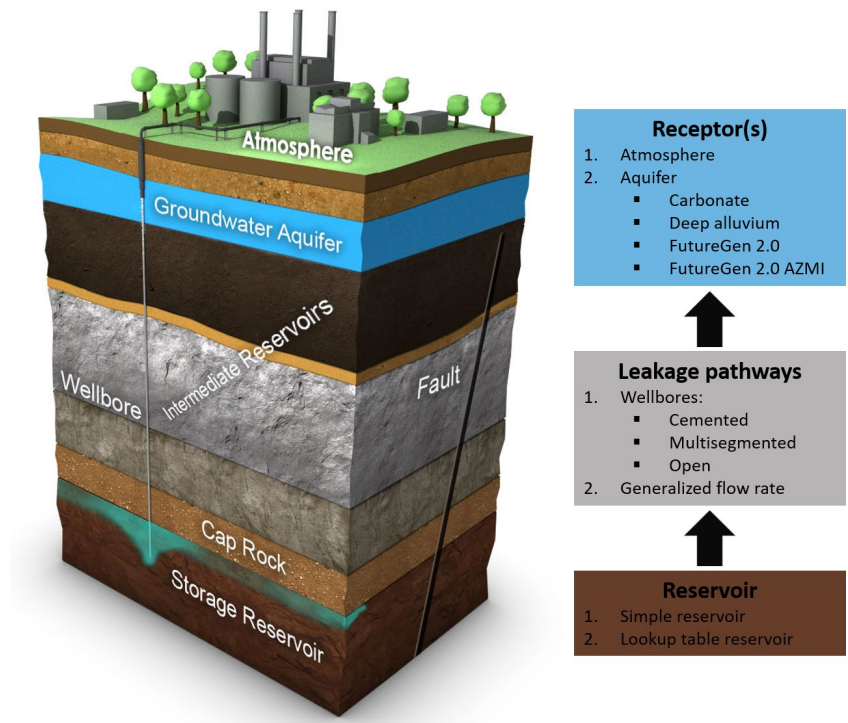

**Figure 2. Base component models of NRAP-Open-IAM (Vasylkivska et al., 2021).**

## 3. Example 1: 3D Synthetic Case

### 3.1. Model description

We first considered a 3D synthetic reservoir model with a $51 \times 51 \times 11$ mesh. The reservoir model is 4 km x 4 km in the horizontal directions. The reservoir is at 1 km depth. Given that this is a synthetic case, we generated monitoring data assuming a ground-truth reservoir model. Figure 3 shows the horizontal log-permeability distribution for the top layer for the ground-truth model and the locations for injection (M3), monitoring (M1, M2, M4 and M5) and legacy wells (L1, L2, …, L5). The remaining 10 layers follow the same horizontal log-permeability distribution as the top layer. The $CO_2$ injection rate is equal to 1 MM tons per year. The injection and post-injection periods are 5 years and 10 years, respectively. The data collected from the monitoring wells and injection well are $CO_2$ saturation and pressure. The data collection frequency is once per month, resulting in 12 measurements per year for $CO_2$ saturation and pressure. The collected data are subsequently assimilated into the prior models to reduce uncertainty in the risk assessment. We generated 100 prior reservoir models using unconditional sequential Gaussian geostatistical simulations. The assimilation of monitoring data to calibrate reservoir models has already been presented in the earlier work of Chen et al. (2020). In this paper, we focus on the dynamic risk assessment using NRAP-Open-IAM based on the simulation results from the calibrated models.

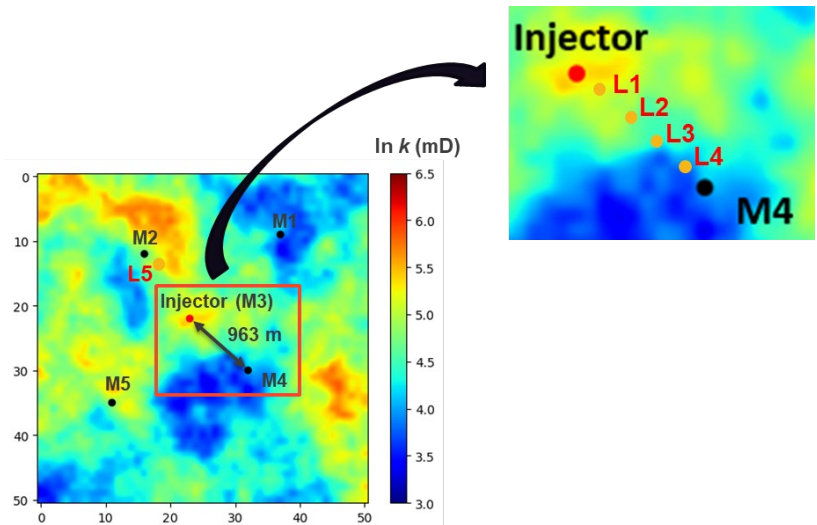

**Figure 3. Horizontal log-permeability distribution for the top layer for the ground-true model and the locations for injector (M3), monitoring wells (M1, M2, M4, and M5) and legacy wells (L1, L2, L3, L4, and L5).**

## 3.2. Results and analysis

### 3.2.1. Effect of monitoring durations

To investigate the impact of monitoring durations on uncertainty reduction in the predictions of risk-related system properties and risk metrics, three different monitoring durations, namely, 0-year, 5-year and 10-year were considered. The legacy well L4 was chosen as the potentially leaky well for the dynamic risk assessment. The predictions of temporal pressure and $CO_2$ saturation in the legacy well L4 and $CO_2$ and brine leakage rates based on the prior models and the updated (posterior) models are presented in Figure 4. As shown in Figure 4, the uncertainties are relatively large in the predictions of pressure, saturation, and $CO_2$/brine leakage rates based on the prior models (0-year monitoring; first column), whereas the uncertainties based on the predictions with updated models after assimilation of 5-year monitoring data are significantly reduced (second column). However, after 5 years (i.e., $CO_2$ injection stops), additional data collection and assimilation does not result in any further reduction in the uncertainty (third column). For instance, the time when $CO_2$ leakage to groundwater aquifer starts ranges from Year 2 to Year 8 based on the predictions with prior models that are not updated with monitoring data. This uncertainty is substantially reduced to a narrower range from 2.8 - 5.1 Years through predictions with the model updated by assimilating 5 years of monitoring data. However, the range does not decrease further through predictions with the updated models by assimilating 10 years of monitoring data. It can also be observed from these figures that $CO_2$ and brine leakage rates remain constant during the post-injection period because the system properties, e.g., pressure and $CO_2$ saturation, reach a steady state, leading to constant leakage rates for $CO_2$ and brine.

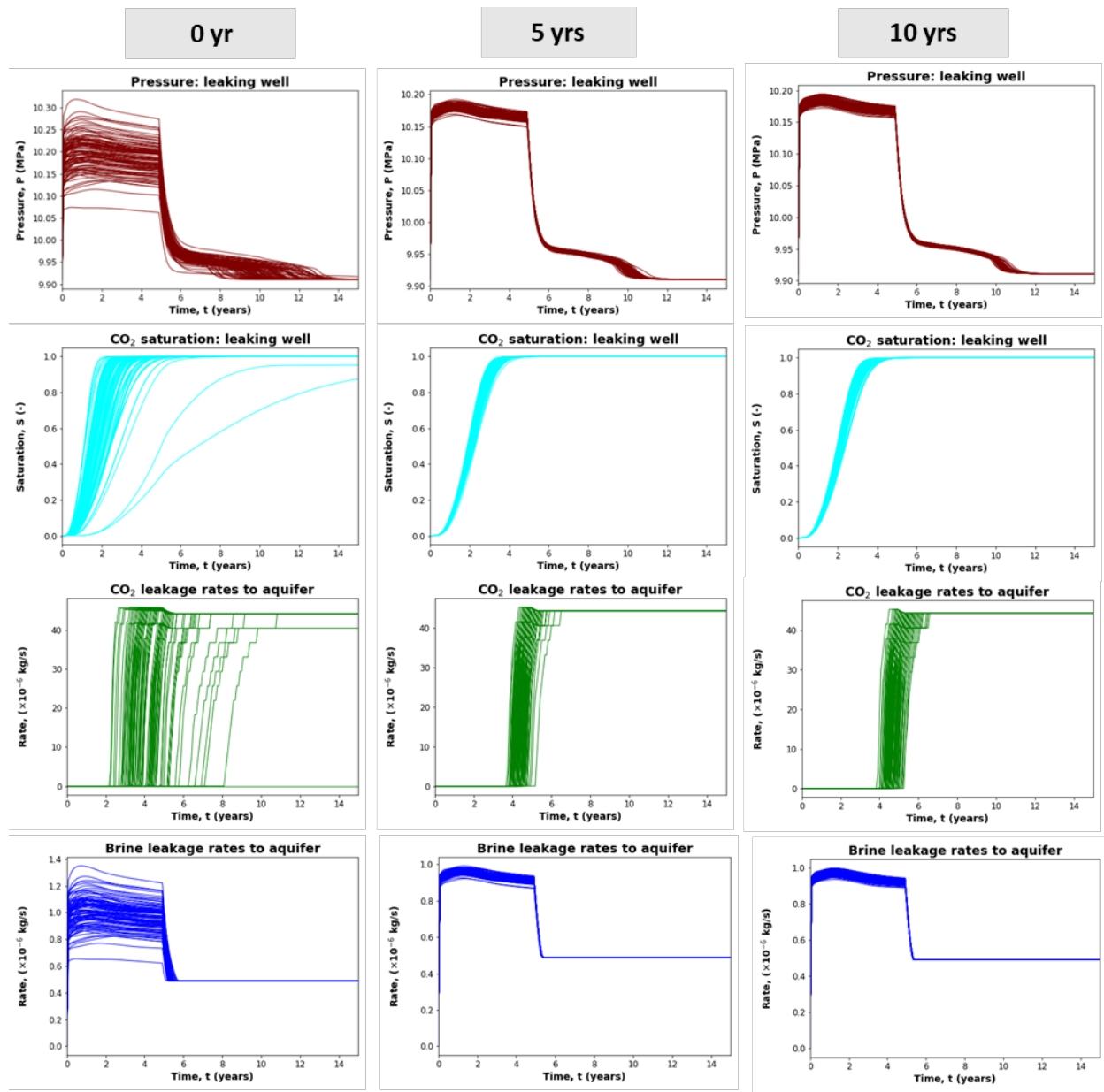

**Figure 4.** The predictions of temporal pressure and $CO_2$ saturation in the legacy well L4 and $CO_2$ and brine leakage rates to aquifer based on the prior models (Column 1) and the updated models with 5 years (Column 2) and 10 years (Column 3) of monitoring data.

### 3.2.2. Effect of legacy well locations

We investigated the effect of legacy well locations. We assumed four different locations between the injector and the monitoring well M4, named L1, L2, L3, and L4 on Figure 3. We considered 5-year monitoring durations and used the updated models with assimilation of 5 years of monitoring data for predictions. The predictions of $CO_2$ and brine leakage rates to aquifer are shown in Figure 5. The figures show that the farther the legacy well is away from the injector, the later $CO_2$ leakage will be observed. Brine leakage is observed in the beginning of injection for all the legacy wells, but the magnitude is different. Note that the uncertainties in the predictions of $CO_2$ and brine leakage are very small. This is because all the predictions were made based on the updated models by assimilating 5 years of monitoring data. Taking L4 as an example, the reason why the observation of brine leakage is earlier than $CO_2$ leakage (see last column in Figure 5) is mainly because it takes a while for the transport of $CO_2$ from injection well to legacy well L4, while it does not take any significant time for brine to be produced from the legacy wells.

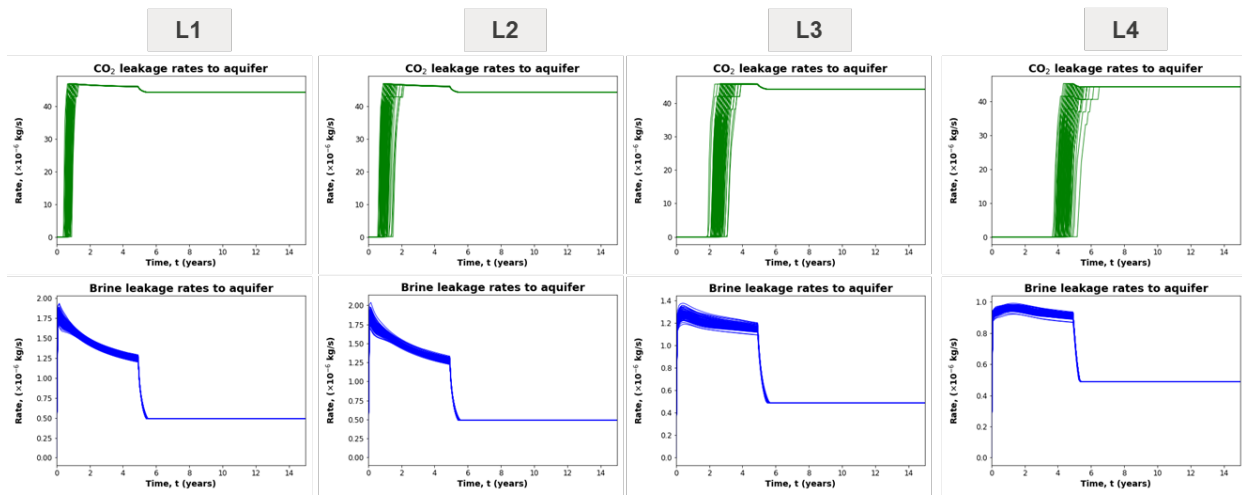

**Figure 5. The predictions of $CO_2$ and brine leakage rates to aquifer at the legacy wells L1, L2, L3, and L4 based on the updated model with 5-year monitoring duration.**

### 3.2.3. Effect of number of monitoring wells

In the work of Chen et al. (2020), we have demonstrated that when more monitoring wells are placed in the reservoir, it usually leads to more significant improvement on reservoir models, i.e., the difference between the calibrated models and the ground-truth model is smaller. In this study, we further investigated the impact of number of monitoring wells on uncertainty reduction in the predictions of risk metrics by sequentially eliminating monitoring wells M1, M5, and M2. M4 remains in all scenarios. For this study, we chose legacy wells L4 and L5 as examples to demonstrate how to evaluate how much information each monitoring well contributes to reducing the uncertainty in leakage rate prediction. As can be observed from Figure 3, the location of the legacy well L4 is close to the monitoring well M4, while the location of the legacy well L5 is close to the monitoring well M2. Figure 6 presents the predictions of brine leakage rates through the legacy wells L4 and L5 based on the updated models with different number of monitoring wells. The first row of the figures corresponds to the predictions based on the updated models with all the monitoring wells (M1, M2, M4, and M5); the second row of the figures corresponds to the predictions based on the updated models with three monitoring wells (M2, M4, and M5); and the third and fourth rows of the figures correspond to the predictions based on the updated models with two (i.e., M2 and M4) and one (i.e., M4) monitoring wells, respectively. As we can see from Figure 6, we do not see any significant uncertainty reduction in the predictions of brine leakage rates when we reduce the monitoring wells M1 and M5 sequentially. This is mainly because both the legacy wells L4 and L5 have an adjacent monitoring well (M4 and M2, respectively) for the first three scenarios (i.e., cases with 4, 3 and 2 monitoring wells). The property (i.e., permeability) around L4 and L5 can be properly updated with the data collected from monitoring wells M4 and M2, so the uncertainty in the predictions such as leakage rates through L4 and L5 can be reasonably small. However, when monitoring well M2 is eliminated and only monitoring well M4 is remaining, only the permeability around well L4 can be properly updated via data assimilation, which can explain why we do not see any significant uncertainty change in the prediction of brine leakage through the legacy well L4, but the uncertainty in the prediction of brine leakage rate through L5 during the injection period (0-5 year) has been significantly increased because no monitoring data are collected around L5. Similar findings have been observed for the prediction of $CO_2$ leakage rates which are not presented in this work. The results show M2 and M4 both contain information to reduce uncertainties in L5 prediction, whereas M4 contains mostly information for the prediction of leakage rate in L4.

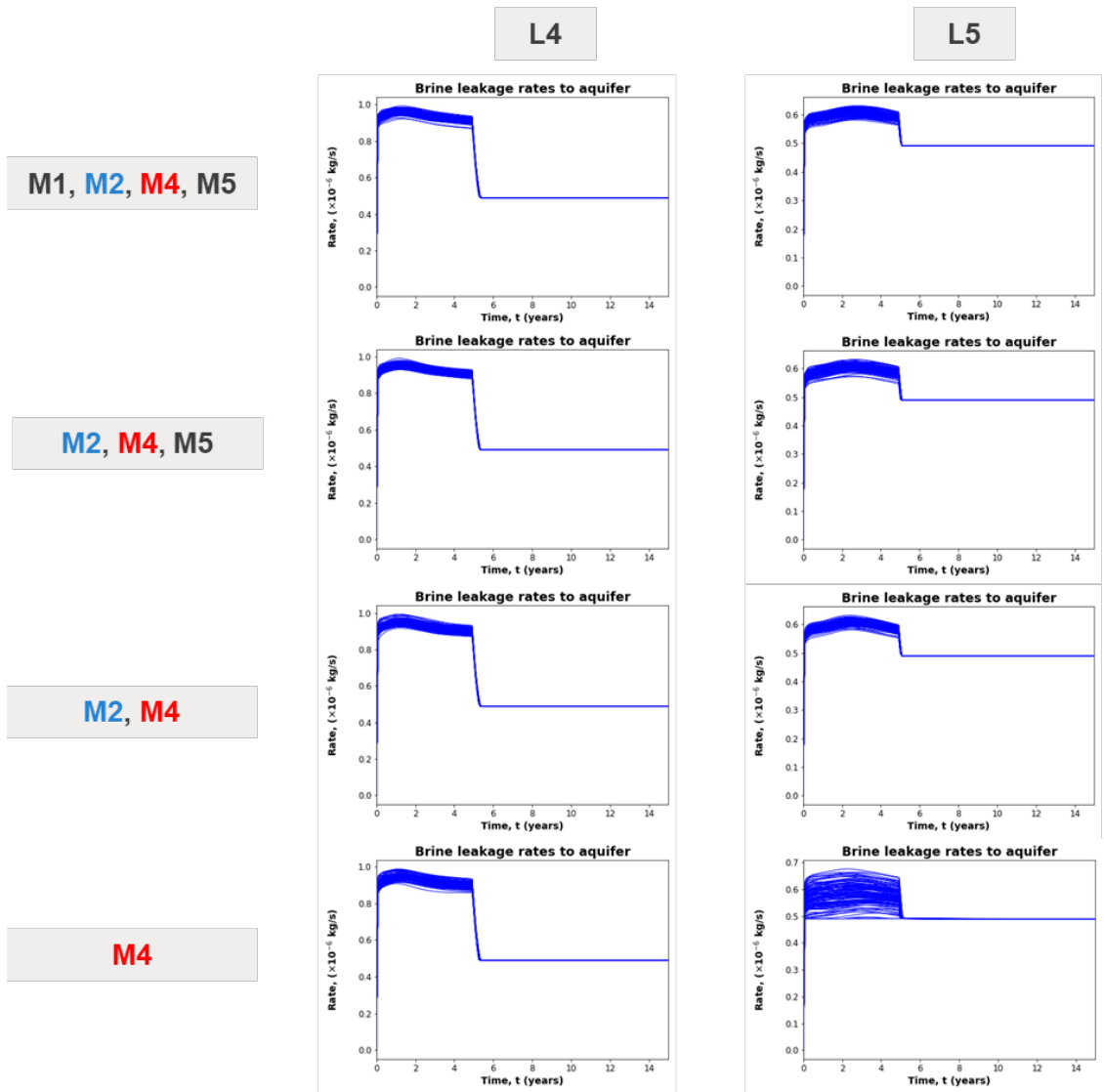

**Figure 6.** The prediction of brine leakage rates through legacy wells L4 and L5 with different number of monitoring wells.

## 4. Example 2: Rock Springs Uplift Storage Site

### 4.1. Site description

The robustness of the proposed framework for dynamic risk assessment was re-evaluated using a hypothetical field site (synthetic case) based on the Rock Springs Uplift (RSU) in Wyoming, USA. The RSU site has been identified as a potential site for geologic $CO_2$ sequestration by the Wyoming Geological Survey (Surdam and Jiao, 2007). Figure 7(a) shows the geologic cross section through the site and the surrounding formations. The Lower Madison formation, as indicated by the red arrow, is one of the target storage reservoirs. The location for the exploratory well RSU 1 is chosen as the location of the $CO_2$ injection well for this study. The model dimension is 6 km by 6 km. The depth of storage reservoir (Lower Madison) ranges from 2.8 km to 4.3 km. The computational mesh of the storage reservoir shown in Figure 7(b) was developed using the Los Alamos Grid Toolbox (George et al., 1999). Figure 7(c) is the permeability distribution for the first layer in the hypothetical ground-truth model. We assume there is one injector (M3, same location as RSU 1) and four monitoring wells (M1, M2, M4, and M5), and here we consider one potential leaking well indicated by the orange dot labeled L1 in Figure 7(c). We consider the scenario where the monitoring measurements are pressures and $CO_2$ saturations. The monitoring data acquisition frequency is once per month. The $CO_2$ injection rate is 1 MM tons per year. We consider 10-year injection and 50-year post-injection periods. As with Example 1, the assimilation of monitoring data to calibrate reservoir models has already been presented in the work of Chen et al. (2020). In this study, we focus on how the risk is dynamically updated using NRAP-Open-IAM based on the simulation results from the calibrated models.

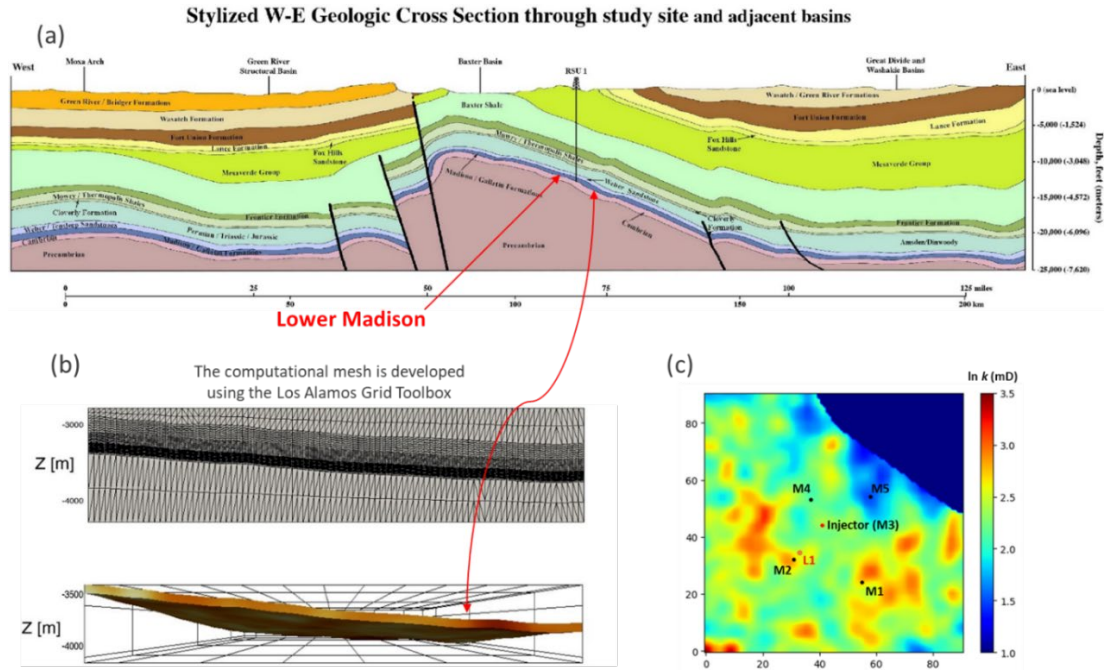

**Figure 7. RSU site description: (a) geologic cross section through the study site and adjacent basins; (b) computational mesh for storage site; (c) ground-truth permeability distribution for the first layer. Geologic cross section in (a) is from Surdam (2013).**

## 4.2. Results and analysis

Figure 8 and Figure 9 show the uncertainty in the predictions of pressure and $CO_2$ saturation plume areas and other risk metrics, respectively, over increasing monitoring durations. As seen in both figures, the predictions made using the models that have not been updated with monitoring data have large uncertainty for both pressure plume area during the injection period (first 10 years) and $CO_2$ saturation plume area during the post-injection period (first columns in Figure 8 and Figure 9). However, with predictions based on updated models after assimilation of three years of monitoring data, the uncertainty in the predictions of these quantities is significantly reduced (second columns in Figure 8 and Figure 9). For the predictions based on the updated models with the assimilation of 15 years of monitoring data, further uncertainty reduction was observed in the predictions for some quantities, e.g., $CO_2$ saturation at legacy well (row 2 column 3 in Figure 9), $CO_2$ leakage rate to groundwater aquifer (row 3 column 3 in Figure 9) and the size of pH plume in groundwater aquifer (last row column 3 in Figure 9), but not for the remaining risk quantities. Note that the pressure plume area reduces to zero during the post-injection period (see first row in Figure 8) because the side boundary condition for the reservoir model is set as a constant pressure boundary. The overpressure in the reservoir from $CO_2$ injection dissipates quickly after $CO_2$ injection stops, which is further demonstrated by the pressure change in the legacy well L1 (see first row in Figure 9). No significant change in the $CO_2$ saturation plume area was observed during the post-injection period because the pressure gradient between the injector and the side boundary substantially decreases after $CO_2$ injection stops and subsequently the movement of $CO_2$ towards the boundary slows down (see second row in Figure 8). The leakages of $CO_2$ and brine to aquifer reach a steady state during the post-injection period because the pressure and $CO_2$ saturation remain constant during the post-injection period. The pH plume volume reaches a pseudo steady state during the post-injection period because of the steady state leakage rate of $CO_2$ from reservoir through wellbore to aquifer (see last row in Figure 9).

It can also be observed that the monitoring data collected during injection period have more value of information than the data collected during post-injection period. The data collected during injection period leads to greater uncertainty reduction in risk-related system properties and risk metrics than the data collected during post-injection period. This is demonstrated with the two examples presented in this paper. In Example 1, most of the uncertainties in risk-related system properties and risk metrics were reduced during the injection period (first 5 years), while in Example 2, we can also see most of the uncertainties were reduced during the injection period (first 3 years). This important observation can guide us when we should stop collecting data from monitoring wells for reducing uncertainty in predictions.

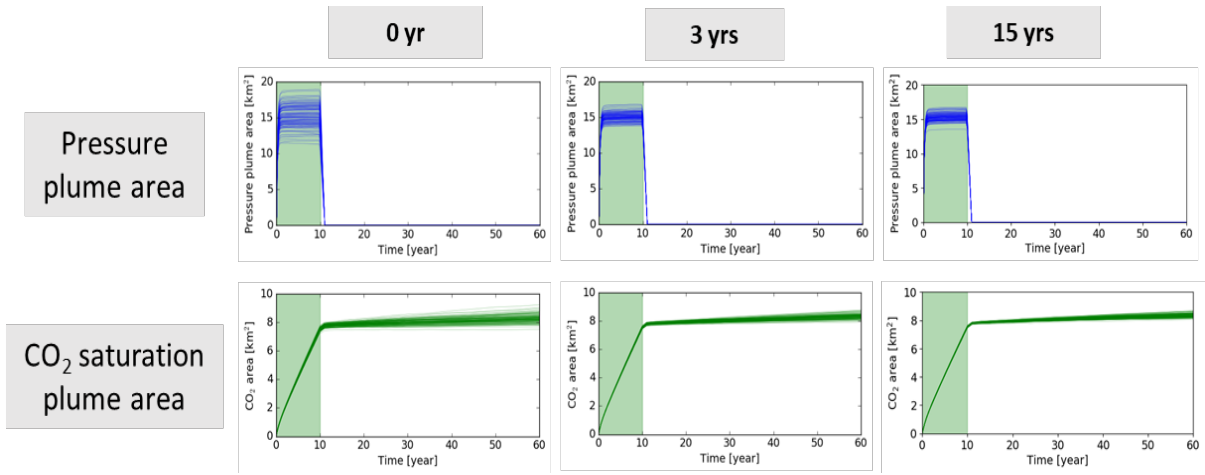

**Figure 8. Uncertainty in pressure/saturation plume areas over monitoring durations, RSU site.**

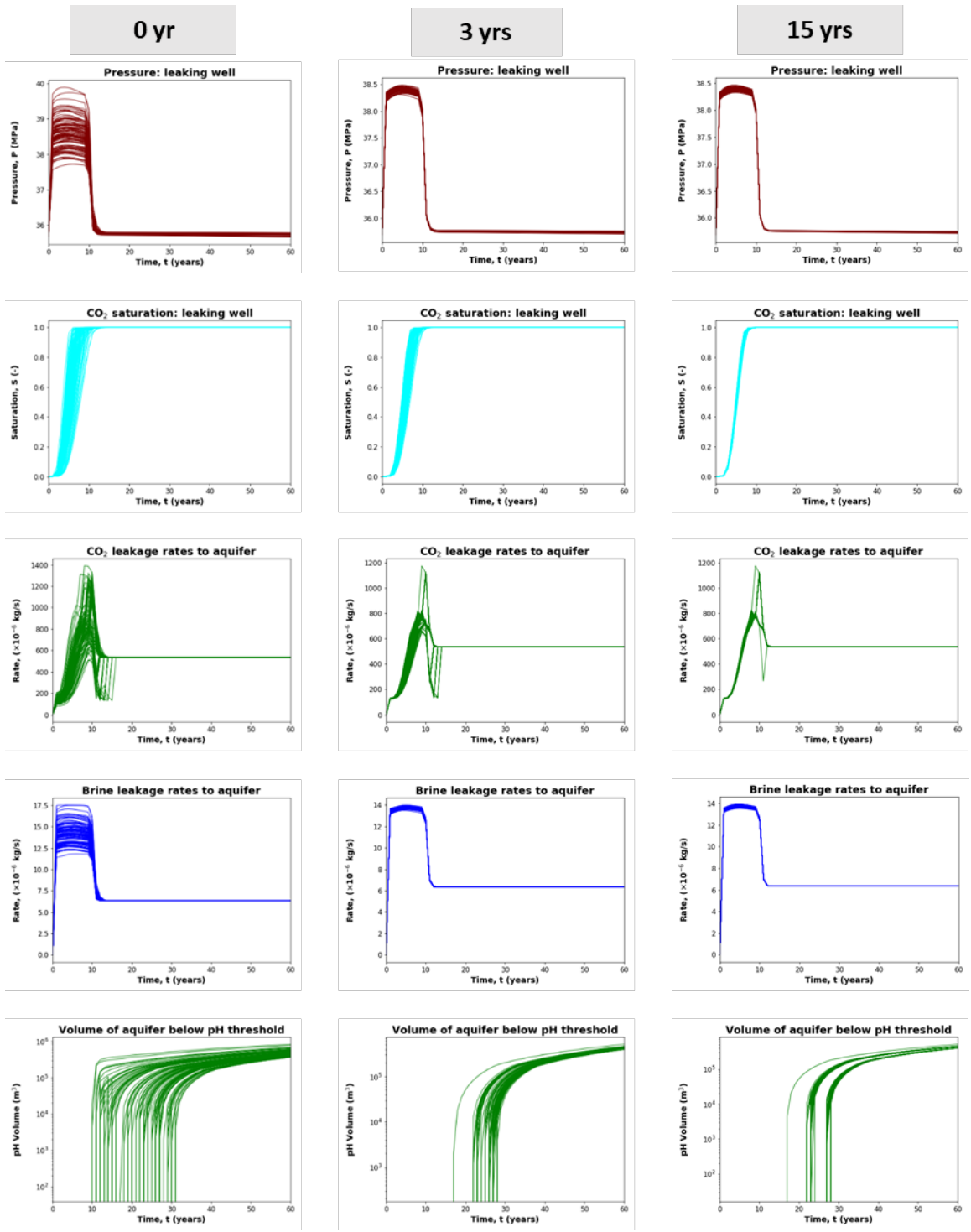

**Figure 9. Uncertainty in risk-related properties and risk metrics over the monitoring duration for the RSU site. The first, second and third columns correspond to uncertainty in risk metrics or quantities based on prior models, calibrated models with 3 years of monitoring data, and calibrated models with 15 years of monitoring data, respectively. "leaking well" in the figures (first two rows) is hypothetical legacy well.**

## 5. Conclusions

We have demonstrated the effectiveness and robustness of the proposed framework based on coupling the conformance evaluation with the NRAP-Open-IAM risk assessment tool for modeling dynamic risk with two case studies: a 3D synthetic example and a synthetic field-scale example based on the Rock Springs Uplift site in Wyoming, USA. The conformance evaluation of a GCS system was performed with a state-of-the-art ensemble-based data assimilation algorithm ES-MDA-GEO. It was observed that ES-MDA-GEO can be utilized to effectively and efficiently assimilate the monitoring measurements collected from $CO_2$ storage operations, and monitoring data assimilation can significantly reduce the uncertainties in predictions of risk-related system properties and risk metrics, e.g., pressure and $CO_2$ saturation plume areas, $CO_2$ and brine leakage rates, groundwater aquifer impact, etc. However, more data or measurements collected from monitoring wells cannot always guarantee more uncertainty reduction in the predictions of these risk-related system properties and risk metrics. We also observed that the monitoring data collected during the injection period have greater value of information than data collected during the post-injection period for uncertainty reduction.

It is important to note that only point data measurements from monitoring wells were considered in the conformance evaluation. This approach is consistent with the capabilities of borehole logging tools for measuring pressure and $CO_2$ saturation, and does not represent a limitation for pressure which tends to spread broadly making point measurements representative of larger-scale averages. On the other hand, for $CO_2$ saturation, it is important to note that local reservoir properties can control saturation and therefore such a local measurement does not provide the kind of integrated or large-scale measurement needed for accurate free-phase $CO_2$ plume delineation. In our future work, we will consider assimilation of spatial measurements, such as $CO_2$ saturation plume interpreted from 4D seismic in the conformance evaluation, and combine them with point measurements for a more comprehensive dynamic risk assessment.

## Acknowledgements

This work was completed as part of the National Risk Assessment Partnership (NRAP) project. Support for this project came from the U.S. Department of Energy's (DOE) Office of Fossil Energy's Coal Research program.

## References

Benson, S. M. and L. Myer (2003). Monitoring to ensure safe and effective geologic sequestration of carbon dioxide. Workshop on carbon dioxide capture and storage.

Chen, B., D. R. Harp, Y. Lin, E. H. Keating and R. J. Pawar (2018). "Geologic $CO_2$ sequestration monitoring design: A machine learning and uncertainty quantification based approach." Applied Energy **225**: 332-345.

Chen, B., D. R. Harp, Z. Lu and R. J. Pawar (2020). "Reducing uncertainty in geologic $CO_2$ sequestration risk assessment by assimilating monitoring data." International Journal of Greenhouse Gas Control **94**(102926).

Condor, J., D. Unatrakarn, M. Wilson and K. Asghari (2011). "A comparative analysis of risk assessment methodologies for the geologic storage of carbon dioxide." Energy Procedia **4**: 4036-4043.

De Lary, L., J. C. Manceau, A. Loschetter, J. Rohmer, O. Bouc, I. Gravaud, C. Chiaberge, P. Willaume and T. Yalamas (2015). "Quantitative risk assessment in the early stages of a $CO_2$ geological storage project: implementation of a practical approach in an uncertain context." Greenhouse Gases: Science and Technology **5**(1): 50-63.

Emerick, A. A. (2018). "Deterministic ensemble smoother with multiple data assimilation as an alternative for history-matching seismic data." Computational Geosciences **22**(5): 1175-1186.

Emerick, A. A. and A. C. Reynolds (2013). "Ensemble smoother with multiple data assimilation." Computers & Geosciences **55**: 3-15.

Evensen, G. (2018). "Analysis of iterative ensemble smoothers for solving inverse problems." Computational Geosciences **22**(3): 885-908.

George, D., A. Kuprat, N. Carlson and C. Gable (1999). "LaGriT–Los Alamos Grid Toolbox."

González-Nicolás, A., A. Cihan, R. Petrusak, Q. Zhou, R. Trautz, D. Riestenberg, M. Godec and J. T. Birkholzer (2019). "Pressure management via brine extraction in geological $CO_2$ storage: Adaptive optimization strategies under poorly characterized reservoir conditions." International Journal of Greenhouse Gas Control **83**: 176-185.

Harp, D. R., R. Pawar, J. W. Carey and C. W. Gable (2016). "Reduced order models of transient $CO_2$ and brine leakage along abandoned wellbores from geologic carbon sequestration reservoirs." International Journal of Greenhouse Gas Control **45**: 150-162.

Kim, S., B. Min, K. Lee and H. Jeong (2018). "Integration of an iterative update of sparse geologic dictionaries with ES-MDA for history matching of channelized reservoirs." Geofluids **2018**.

Le, D. H., A. A. Emerick and A. C. Reynolds (2016). "An adaptive ensemble smoother with multiple data assimilation for assisted history matching." SPE Journal **21**(2): 195–207.

Li, Q. and G. Liu (2016). Risk assessment of the geological storage of $CO_2$: A review. Geologic Carbon Sequestration, Springer**:** 249-284.

Nicot, J.-P., C. M. Oldenburg, J. E. Houseworth and J.-W. Choi (2013). "Analysis of potential leakage pathways at the Cranfield, MS, USA, $CO_2$ sequestration site." International Journal of Greenhouse Gas Control **18**: 388-400.

Oladyshkin, S., H. Class and W. Nowak (2013). "Bayesian updating via bootstrap filtering combined with data-driven polynomial chaos expansions: methodology and application to history matching for carbon dioxide storage in geological formations." Computational Geosciences **17**(4): 671-687.

Oldenburg, C. M. (2018). "Are we all in concordance with the meaning of the word conformance, and is our definition in conformity with standard definitions?" Greenhouse Gases: Science and Technology **8**(2): 210-214.

Onishi, T., M. C. Nguyen, J. W. Carey, B. Will, W. Zaluski, D. W. Bowen, B. C. Devault, A. Duguid, Q. Zhou, S. H. Fairweather, L. H. Spangler and P. H. Stauffer (2019). "Potential $CO_2$ and brine leakage through wellbore pathways for geologic $CO_2$ sequestration using the National Risk Assessment Partnership tools: Application to the Big Sky Regional Partnership." International Journal of Greenhouse Gas Control **81**: 44-65.

Pawar, R., G. Bromhal, R. Dilmore, B. Foxall, E. Jones, C. Oldenburg, P. Stauffer, S. Unwin and G. Guthrie (2013). Quantification of Risk Profiles and Impacts of Uncertainties as part of US DOE's National Risk Assessment Partnership (NRAP). GHGT-11.

Pawar, R., G. S. Bromhal, S. Chu, R. M. Dilmore, C. M. Oldenburg, P. H. Stauffer, Y. Zhang and G. D. Guthrie (2016). "The National Risk Assessment Partnership's integrated assessment model
26

for carbon storage: A tool to support decision making amidst uncertainty." International Journal of Greenhouse Gas Control **52**: 175-189.

Rafiee, J. and A. C. Reynolds (2017). "Theoretical and efficient practical procedures for the generation of inflation factors for ES-MDA." Inverse Problems **33**(115003).

Rafiee, J. and A. C. Reynolds (2017). "Theoretical and efficient practical procedures for the generation of inflation factors for ES-MDA." Inverse Problems **33**: 115003.

Sambandam, S. T. (2018). Optimization of $CO_2$ storage systems with constrained bottom-hole pressure injection, Master's Thesis, Stanford University, Department of Energy Resources Engineering.

Silva, V. L. S., A. A. Emerick, P. Couto and J. L. D. Alves (2017). "History matching and production optimization under uncertainties–Application of closed-loop reservoir management." Journal of Petroleum Science and Engineering **157**: 860-874.

Stauffer, P. H., H. S. Viswanathan, R. J. Pawar and G. D. Guthrie (2009). "A System Model for Geologic Sequestration of Carbon Dioxide." Environmental Science & Technology **43**(3): 565-570.

Sun, W. and L. J. Durlofsky (2019). "Data-space approaches for uncertainty quantification of $CO_2$ plume location in geological carbon storage." Advances in Water Resources **123**: 234-255.

Surdam, R. C. (2013). Geological $CO_2$ storage characterization: The key to deploying clean fossil energy technology, Springer Science & Business Media.

Surdam, R. C. and Z. Jiao (2007). The Rock Springs Uplift: An outstanding geological $CO_2$ sequestration site in southwest Wyoming, Wyoming State Geological Survey.

Vasylkivska, V., R. Dilmore, G. Lackey, Y. Zhang, S. King, D. Bacon, B. Chen, K. Mansoor and D. Harp (2021). "NRAP-Open-IAM: A Flexible Open Source Integrated Assessment Model for Geologic Carbon Storage Risk Assessment and Management " Environmental Modelling & Software **In review**.

Xiao, T., B. McPherson, R. Esser, W. Jia, Z. Dai, S. Chu, F. Pan and H. Viswanathan (2020). "Chemical Impacts of Potential $CO_2$ and Brine Leakage on Ground water Quality with Quantitative Risk Assessment: A Case Study of the Farnsworth Unit." Energies **13**(24): 6574.

Yonkofski, C. M., J. A. Gastelum, E. A. Porter, L. R. Rodriguez, D. H. Bacon and C. F. Brown (2016). "An optimization approach to design monitoring schemes for $CO_2$ leakage detection." International Journal of Greenhouse Gas Control **47**: 233-239.

Zhang, Q., S. Jiang, X. Wu, Y. Wang and Q. Meng (2020). "Development and Calibration of a Semianalytic Model for Shale Wells with Nonuniform Distribution of Induced Fractures Based on ES-MDA Method." Energies **13**(14): 3718.

Zhang, Y., P. Vouzis and N. V. Sahinidis (2011). "GPU simulations for risk assessment in $CO_2$ geologic sequestration." Computers & Chemical Engineering **35**(8): 1631-1644.

Zhang, Z. and R. Agarwal (2013). "Numerical simulation and optimization of $CO_2$ sequestration in saline aquifers." Computers & Fluids **80**: 79-87.

Zhao, Y., F. Forouzanfar and A. C. Reynolds (2017). "History matching of multi-facies channelized reservoirs using ES-MDA with common basis DCT." Computational Geosciences **21**(5): 1343-1364.